# Time, Incompleteness and Singularity in Quantum Cosmology

Philip V. Fellman[1], Jonathan Vos Post[2], Christine Carmichael[3], Alexandru Manus[1], and Dawna Lee Attig[1]

[1] Southern New Hampshire University, Manchester, NH, USA
`Shirogitsune99@yahoo.com, alexandru.manus@snhu.edu,`
`adawnalee@msn.com`
[2] Computer Futures, Inc., Altadena, CA, USA
`jvospost3@gmail.com`
[3] Woodbury University, Burbank, CA USA
`cmcarmichael@yahoo.com`

**Abstract.** In this paper we extend our 2007 paper, "Comparative Quantum Cosmology: Causality, Singularity, and Boundary Conditions", http://arxiv.org/ftp/arxiv/papers/0710/0710.5046.pdf, to include consideration of universal expansion, various implications of extendibility and incompleteness in spacetime metrics and, absent the treatment of Feynman diagrams, the use of Penning trap dynamics to describe the Hamiltonians of space-times with no characteristic upper or lower bound.

## 1 Introduction

Cosmology, and more particularly, quantum cosmology, has generated a number of different, often radically different viewpoints about both the beginning and the ending of the universe as well as a variety of possibilities with respect to cyclical theories of expansion and contraction. Despite the manifold differences between these theories, many of them, in particular, the theories expounded by Hawking (1999), Carroll and Chen (2004, 2005) and Peter Lynds (2003, 2006) often have more elements in common than those among which they differ. In particular, the commonalities tend to cluster around the concepts of "weak singularity", "no boundary condition", and "the problem of specialness" with respect to scale, entropy and initial conditions.

In the present paper we examine these concepts with particular attention to the nature and role of singularities relying in large part on two particular sources for our analysis. The first is the 1994 work of C.J.S. Clark on the analysis of space-time singularities, a work of general relativity largely exclusive of quantum mechanical effects and the second is a very recent paper by David J. Fernandez C. and Mercedes

Velazquez (2008) whose mathematics we employ to explore at a quantum mechanical level the turning points of cosmic expansion (i.e. maximal spaces with no further possible extensions) and the "big bang/big crunch" particularly in the context of the limitations imposed on the order of events suggested by Lynds (2006).

### 1.1 The Mechanics of Contraction – The Standard Approach

"Big Crunch" theories all depend on some sort of universal contraction although not nearly enough of these theories offer a plausible mechanism for why the universe should suddenly contract at a particular point. That is, "by what mechanism should the universe cease expanding, at which point gravity initiates an apparently unstoppable contraction?" The problem is sufficiently deep that for many years most physicists assumed a "steady state" model of the universe. In his treatment of incomplete geodesics and maximal spaces, C.J.S. Clarke offers at least one possibility in his treatment of the maximality assumption (p. 8):

> The foregoing result has shown that if M is inextendible then there is some timelike curve (actually a geodesic) - i.e. a possible worldline of a particle - which could continue in some extension of M but which in M itself simply stops. This seems unreasonable: why should M be cut short this way? It seems natural to demand that "if a space-time can continue then it will"; in other words, to demand that any reasonable space-time should not be inextendible. This is an assumption imposed upon space-time in addition to the field equations of Einstein.
>
> It can easily be shown that any space-time can in fact be extended until no further extension is possible. At this point the space-time is called maximal, and so we are lead to the idea that we need only consider maximal space-times. But this idea is really not as innocuous as it might seem because of the problem that the extension of a space-time, when it exists, cannot usually be determined uniquely. In special cases there are unique extensions:
> an analytic space-time has (subject to some conditions) a unique maximal analytic extension; similarly, a global hyperbolic solution of the field equations (with a specified level of differentiability) is contained in a unique maximal solution. In both these cases a "principle of sufficient reason" demands that the maximal solution be taken. But suppose one has a non-analytic space-time where Einstein's field equations fail to predict a unique extension (either because there is a Cauchy horizon or because there is some sort of failure of the differentiability needed for the existence of unique solutions). Or suppose a situation arises in which there is a set of incomplete curves, each one of which can be extended in some extension of the space-time, but where there is no extension in which they can all be extended. (There exists admittedly artificial examples of this (Misner, 1967). In cases such as these, the same principle of sufficient reason would not allow one extension to exist at the expense of another. Perhaps the space-time, like Buridan's ass between two bales of hay, unable to decide which way to go, brings the whole of history to a halt.

## 2 The Quantum Mechanical Approach to Maximal States

In at least one sense we must "jump the gun" a bit in order to discuss the quantum mechanical argument for the maximal extension of space-time never being achieved because this result is merely a byproduct of a quantum mechanical argument against the existence of strong singularity. The gist of Fernandez and Velazques proof is that (8.0) "in identifying the appropriate displacement operator as well as an adequate 'extremal' state for the Penning trap cavity…the corresponding Hamiltonian has neither a ground nor a top energy eigenvalue. The proof of this claim, which we will return to in our discussion of singularities is as follows and is drawn directly from the previously reference paper by Fernandez and Velazquez. In order to understand the proof, we must begin with their demonstration of the extremal state wave function:

This existence of the extremal state $|0,0,0\rangle$ is guaranteed by a theorem which is proven elsewhere [5]. It ensures that, if the operators

$$B_j = i\vec{P} \cdot \vec{\alpha} + \vec{R} \cdot \vec{\beta}_j, \quad B_j^\dagger = -i\vec{\alpha}_j^\dagger \cdot \vec{P} + \vec{\beta}_j^\dagger \cdot \vec{R}, \quad j = 1,2,3, \qquad (1)$$

Obey the commutation relations (17), then the system of partial differential equations $\langle \vec{r}|B_j|0,0,0\rangle\ 0, j = 1,2,3$, for the extremal state wave function $\phi_0(\vec{r}) \equiv \langle \vec{r}|0,0,0\rangle$ has a square integrable solution given by

$$\phi_0(\vec{r}) = c \exp\left(-\tfrac{1}{2} a_{ij} x_i x_j\right) = c \exp\left(-\tfrac{1}{2} \vec{r}^T \mathbf{a} \vec{r}\right), \qquad (2)$$

Where $\mathbf{a} = (a_{ij})$ is a complex symmetric matrix satisfying

$$\mathbf{a}\vec{\alpha}_j = \vec{\beta}_j, \quad j = 1,2,3, \qquad (3)$$

According to (22), through equations (12.16) we identify the vectors

$$\vec{\alpha}_1 = \tfrac{1}{2(b^2+v)^{1/4}} (1,-i,0)^T, \quad \vec{\beta}_1 = (b^2+v)^{1/2}\vec{\alpha}_1,$$

$$\vec{\alpha}_2 = -\tfrac{1}{2(b^2+v)^{1/4}} (1,i,0)^T, \quad \vec{\beta}_2 = (b^2+v)^{1/2}\vec{\alpha}_2, \qquad (4)$$

$$\vec{\alpha}_3 = -\tfrac{1}{\sqrt{2}(-2v)^{1/4}} (0,0,1)^T, \quad \vec{\beta}_3 = (-2v)^{1/2}\vec{\alpha}_3,$$

Thus, $\mathbf{a} = \mathrm{diag}\left[\sqrt{b^2+v}, \sqrt{b^2+v}, \sqrt{-2v},\right]$, and from (23) we finally get the extremal state wave function we were looking for:

$$\phi_0(\vec{r}) = c \exp\left(-\tfrac{\sqrt{b^2+v}}{2}(x^2+y^2) - \sqrt{\tfrac{-v}{2}}z^2\right). \qquad (5)$$

What is particularly interesting about this extremal state is that the Hamiltonian has neither a ground nor a top energy eigenvalue (p. 9) This is given a further exposition in the section on mean values of physical quantities. In particular, this calculation again demonstrates that the values of the extremal state as well as the CS wave function coincide with the previous calculation. Mathematically this is shown by the following exposition (pp. 7-8):

Let us evaluate next the mean values $\langle X_j \rangle_z \equiv \langle z|X_j|z\rangle, \langle P_j\rangle_z \equiv \langle z|P_j|z\rangle, j = 1,2,3$, and the corresponding mean square deviations in a given CS$|z\rangle$. To do that, we analyze first how the operators $X_j, X_j^2, P_j, P_j^2$ are transformed under D(z). By using equation (35) it is straightforward to show that:

$$D^\dagger(z) X_j^\eta D(z) = (X_j + \Gamma_j)^\eta. \quad D^\dagger(z) P_j^\eta D(z) = (P_j + \Sigma_j)^\eta, \eta = 1,2 \ldots \quad (6)$$

Therefore:

$$\langle X_j \rangle_z = \langle X_j \rangle_0 + \Gamma_j, \langle X_j^2 \rangle_z = \langle X_j^2 \rangle_0 + 2\Gamma_j \langle X_j \rangle_0 + \Gamma_j^2, (\Delta X_j)_z^2 = (\Delta X_j)_0^2 \quad (7)$$

$$\langle P_j \rangle_z = \langle P_j \rangle_0 + \Sigma_j, \langle P_j^2 \rangle_z = \langle P_j^2 \rangle_0 + 2\Sigma_j \langle P_j \rangle_0 + \Sigma_j^2, (\Delta P_j)_z^2 = (\Delta P)_0^2 \quad (8)$$

Notice that the mean square deviations of $X_j$ and $P_j$ are independent of $z_1, z_2, z_3$ but depend on $\langle X_j \rangle_0, \langle P_j \rangle_0, \langle X_j^2 \rangle_0, \langle P_j^2 \rangle_0, j = 1,2,3$, which need to be evaluated. The first six quantities can be obtained from the homogeneous equations $\langle B_k \rangle_0 = i(\vec{\alpha}_k)_j \langle P_j \rangle_0 + (\vec{\beta}_k)_j \langle X_j \rangle_0 = 0, \langle B_k^\dagger \rangle_0 = -i(\vec{\alpha}_k^*)_j \langle P_j \rangle_0 + (\vec{\beta}_k^*)_j \langle X_j \rangle_0 = 0, k = 1,2,3$ (see (22)) and use that $B_k |0,0,0\rangle = \langle 0,0,0| B_k^\dagger = 0$. By using (25), the system to be solved becomes:

$$-i\sqrt{-2v}\, \langle Z \rangle_0 + \langle P_z \rangle_0 = 0$$

$$\sqrt{b^2 + v}(\langle X \rangle_0 - i Y_0) + i(\langle P_x \rangle_0 - i\langle P_y \rangle_0) = 0,$$

$$-\sqrt{b^2 + v}(\langle X \rangle_0 + i Y_0) - i(\langle P_x \rangle_0 + i\langle P_y \rangle_0) = 0,$$

and the complex conjugate equations. Its solution is given by

$$\langle X_j \rangle_0 = \langle P_j \rangle_0 = 0, \quad j = 1,2,3. \quad (41)$$

In order to obtain $\langle X_j^2 \rangle_0, \langle P_j^2 \rangle_0$, we calculate the mean values for the several products of pairs involving $B_j, B_k^\dagger$. From these thirty six equations just twenty one are linearly independent: $\langle B_j B_k \rangle_0 = 0, j = 1,2,3, k \leq j$ (six equations); $\langle B_j^\dagger B_k^\dagger \rangle_0 = 0, j = 1,2,3, k \leq j$ (six equations); $\langle B_k^\dagger B_j \rangle_0 = 0, j,k = 1,2,3,$ (nine equations). By solving this linear system, the non-null results for the mean values of the twenty one independent products of $X_i$ and $P_j$ are now:

$$\langle X^2 \rangle_0 = \langle Y^2 \rangle_0 = [4(b^2 + v)]^{-\frac{1}{2}}, \quad \langle Z^2 \rangle_0 = (-8v)^{-\frac{1}{2}},$$

$$\langle P_x^2 \rangle_0 = \langle P_y^2 \rangle_0 = [(b^2 + v)/4]^{\frac{1}{2}}, \quad \langle P_z^2 \rangle_0 = (-v/2)^{\frac{1}{2}},$$

$$\langle X P_x \rangle_0 = \langle Y P_y \rangle_0 = \langle Z P_z \rangle_0 = i/2.$$

The previous formulas imply that equations (39, 40) become

$$(\Delta X)^2_z = (\Delta Y)^2_z = [4(b^2 + v)]^{-\frac{1}{2}}, \quad (\Delta Z)^2_z = (-8v)^{-\frac{1}{2}},$$

$$(\Delta P_x)^2_z = (\Delta P_y)^2_z = [(b^2 + v)/4]^{\frac{1}{2}}, \quad (\Delta P_z)^2_z = (-v/2)^{\frac{1}{2}},$$

and therefore

$$(\Delta X)_z(\Delta P_x)_z = (\Delta Y)_z(\Delta P_y)_z = (\Delta Z)_z(\Delta P_z)_z = 1/2.$$

This means that our CS have minimum Heisenberg uncertainty relations.

Finally, by using equations (15, 21) we calculate the mean value of the Hamiltonian $H$ in a given CS $|z\rangle$:

$$\langle H \rangle_z = \omega_1|z_1|^2 - \omega_2|z_2|^2 + \omega_3|z_3|^2 + E_{0,0,0}. \tag{42}$$

A similar calculation for $(H^2)_z$ can be done, leading to:

$$(\Delta H)^2_z = \left(b + \sqrt{b^2 + v}\right)^2 |z_1|^2 + \left(b - \sqrt{b^2 + v}\right)^2 |z_2|^2 - 2v|z_3|^2. \tag{43}$$

Once again, the fact that $H$ is not positive definite is clearly reflected in (42).

Along this work we have assumed that $b = -\frac{eB}{2c} > 0$. For $b < 0$, small differences concerning the identification of the appropriate annihilation and creation operators arise. However, the extremal state and CS wave functions $\phi_0(\vec{r}), \phi_z(\vec{r})$ as well as the corresponding mean values, will coincide with those previously calculated. In particular, the Heisenberg uncertainty relation will achieve once again its minimum value [14].

## 3 Singularity – Early Arguments

One of the simplest arguments against strong singularity is simply that of a choice of improper metric. In this case, an inappropriate choice of the Schwarzschild metric as a representation of strong or true singularity. As Clarke explains:

> "In 1924 Eddington showed there was an isometry between the space-time M defined by the region r>2m in the Schwarzschild metric and part of a larger space-time M'. Incomplete curves in M on which r -> 2m were mapped by this isometry into curves which were extensible in M': The singularity at 2m was no longer present. So if we identify the Schwarzschild space-time with the part of the Eddington space-time M' with which it is isometric, we see that it is not just incomplete in the formal sense defined above: it actually had a piece missing from it, a piece that is restored at M'. The singularity at r=2m is thus a mathematical artifact, a consequence of the fact that the procedure used to solve the field equations had fortuitously produced only a part of the complete space.
> We note that, despite this, there are still some authors who regard the Schwarzschild 'singularity' at r=2m as genuine; but this is only

> justified if (as done by Rosen, 1974) one uses a non-standard physical theory in which there is some additional structure (such as a background metric) which itself becomes singular under the under the isometry of the metric into M', so that one structure, the metric, or the background is always singular at r = 2m.
>
> The situation in Schwarzschild clearly contrasts with that of the Friedmann metrics. For these, on any of the incomplete curves the Ricci scalar tends to infinity. For the smooth space-times that we are considering at the moment this is impossible on a curve which has an endpoint in space-time, and so there can in this case, be no isometric M' in which these curves have an endpoint."

Clarke's subsequent arguments rest on the properties of extensibility and global hyperbolicity[1] and ends up with various cosmic censorship models (both strong and weak forms) which either prevent the existence of strong singularities or which make them inaccessible to any particular observer in an inertial frame with bounded acceleration. Much of the distinction between these theories is a function of globally hyperbolic space-time and past-simplicity/past hyperbolicity such that for all timelike geodesics k in $U_0(s)$, $\mathcal{J}(k) < 2$; also geodesics in $U_0(s)$ have no conjugate points (p. 127) From here he proceeds to eliminate what her refers to "primal singularities" and "dragging geodesics". In this context, cosmic censorship provides a number of mechanisms for explaining why strong singularities are not accessible. However, Peter Lynds offers a rather different approach, based on the second law of thermodynamics suggesting that the very nature of the big bang-big crunch makes any strong singularity inaccessible, even if it exists within the light cone and that this inaccessible history also explains the problems of scale and naturalness raised by Carroll and Chen (2004, 2005).

## 4  Time

In 2003, Peter Lynds published a controversial paper, "Time and Classical and Quantum Mechanics: Indeterminacy vs. Discontinuity" in Foundations of Physics Letters. Lynds' theory does away with the notion of "instants" of time, relegates the "flow of time" to the psychological domain. While we have commented elsewhere on the implications of Lynds' theory for mathematical modeling it might save a bit of time simply to borrow Wikipedia's summary of this paper:

> Lynds' work involves the subject of time. The main conclusion of his paper is that there is a necessary trade off of all precise physical magnitudes at a time, for their continuity over time. More specifically, that there is not an instant in time underlying an object's motion, and as its position is constantly changing over time, and as such, never determined, it also does not have a determined relative position. Lynds posits that this is also the correct resolution of Zeno's paradoxes, with the paradoxes arising because people have wrongly assumed that an object in motion has a determined relative position at any given instant in time, thus rendering the body's motion static and frozen at that instant and enabling the impossible situation of the paradoxes to be derived. A further implication of this conclusion is that if there is no such thing as determined relative position, velocity, acceleration, momentum, mass, energy and all other physical magnitudes, cannot be precisely determined at any time either. Other implications of Lynds' work are that time does not flow, that in relation to indeterminacy in precise physical magnitude, the micro and macroscopic are inextricably linked and both a part of the same parcel, rather than just a case of the former underlying and contributing to the latter, that Chronons, proposed atoms of time, cannot exist, that it does not appear necessary for time to emerge or congeal from the big bang, and that Stephen Hawking's theory of Imaginary time would appear to be meaningless, as it is the relative order of events that is relevant, not the direction of time itself, because time does not go in any direction. Consequently, it is meaningless for the order of a sequence of events to be imaginary, or at right angles, relative to another order of events.

---

[1] For every pair of points, $p, q \in U$, $I^-(p) \cap I^+(q)$ is compact. Here $I^{\pm}$ is the future(past of a set $S$ in space-time. "Causality" holds on $U$ (no closed timelike curves exist). Classically, a more restrictive and technical assumption is required, namely, "strong causality" – that no "almost closed" timelike curves exist, but the recent work of Hawking and Penrose (1996) shows that causality suffices. Global hyperbolicity implies that there is a family of Cauchy surfaces for $U$. Essentially, it means that everything that happens on $U$ is determined by the equations of motion, together with initial data specified on a surface. (Wikipedia).

One can see from the above summary that this radical reformulation of the concept of time is bound to have significant cosmological implications. We discussed some of these implications in a brief paper in 2004, "Time and Classical and Quantum Mechanics and the Arrow of Time".[2] We began with a discussion of John Gribbin's analysis, "Quantum Time Waits for No Cosmos", and Gribbin's argument against the mechanics of time reversibility as an explanation for the origin of the universe, where he cites Raymond LaFlamme:[3]

> The intriguing notion that time might run backwards when the Universe collapses has run into difficulties. Raymond LaFlamme, of the Los Alamos National Laboratory in New Mexico, has carried out a new calculation which suggests that the Universe cannot start out uniform, go through a cycle of expansion and collapse, and end up in a uniform state. It could start out disordered, expand, and then collapse back into disorder. But, since the COBE data show that our Universe was born in a

---

[2] "Time and Classical and Quantum Mechanics and the Arrow of Time", paper presented at the annual meeting of the North American Association for Computation in the Social and Organizational Sciences, Carnegie Mellon University, June, 2004.

[3] http://www.lifesci.sussex.ac.uk/home/John_Gribbin/timetrav.htm\

> smooth and uniform state, this symmetric possibility cannot be
> applied to the real Universe.

The concept of time reversibility, while apparently quite straightforward in many cases, seems never to be without considerable difficulty in cosmology, and indeed, explaining the mechanics of time reversibility and its relationship to Einstein's cosmological constant is one of the major enterprises of quantum cosmology (Sorkin, 2007). The terns of the debate expressed by Gribbin above, have been extended by Wald (2005) who acknowledges Carroll and Chen's work, discussed earlier in this paper, but who argues that at some point an anthropic principle introduces a circularity into the causal logic. Sorkin covers much of the distance needed for a Carroll-Chen type counterargument in his 2007 paper, "Is the cosmological "constant" a nonlocal quantum residue of discreteness of the causal set type?", however a complete discussion of Sorkin's model is beyond the scope of our present exploration.[4]

### 4.1 The Arrow of Time

One set of problems with the thermodynamic arrow of time for the very early universe (and this is partially addressed by Sorkin) is that in addition to the problem of possible inhomogeneities in the early universe there is still a lack of consensus or unequivocal evidence on the invariance of fundamental physical constants during the early history of the universe. Various authors have recently suggested that the speed of light may have been greater during the earliest period of the universe's formation (Murphy, Webb and Flambaum, 2002).

In our 2004 review of Lynds work on how we might better understand the thermodynamic arrow of time we also raised two moderately troubling complexity issues which at present remain largely unanswered. The first is the problem of heteroskedastic time behavior in the early universe. This question is not unconnected to the question of changes in the values of fundamental physical constants in the early universe. In most models of early universe formation a smooth or linear flow of time is assumed. However, it is possible to imagine inflationary models where the expansion of time dimension or the time-like dimensions of a higher order manifold inflate in a heteroskedastic fashion. To the extent that the thermodynamic arrow of time is invoked as an element of cosmological explanation, It would need to be able to explain the dynamical evolution of the universe, not just as we know it today, but at those particularly difficult to characterize beginning and end points of the system. The difficulty with heteroskedastic time distributions is that they may or may not allow recovery of the standard Boltzmann expression. At a deeper level, it is likely that in characterizing the development of the early universe, one may have to incorporate a significant number of non-commuting quantum operators. Further, in this context, our knowledge of the early universe is both substantively incomplete, because we lack any system of measurement for the first three hundred thousand years of time evolution of the system (i.e. the period prior to the decoupling of baryons and photons) and very likely theoretically incomplete as well. We can compare the problem to one of discrete time series evolutions with low dimensionality and discrete combinatorics. For example, the random order of a shuffled deck of cards can eventually be repeated because the dimensionality of the system is low. As dynamical systems take on higher orders of dimensionality their asymptotes become ill-defined (in at least one sense, this is the objection raised by Gribbin and LaFlamme).[5]

---

[4] In his conclusion, Sorkin argues "Heuristic reasoning rooted in the basic hypotheses of causal set theory predicted $\Lambda \sim \pm 1/\sqrt{V}$, in agreement with current data. But a fuller understanding of this prediction awaits the "new QCD" ("quantum causet dynamics"). Meanwhile, a reasonably coherent phenomenological model exists, based on simple general arguments. It is broadly consistent with observations but a fuller comparison is needed. It solves the "why now" problem: $\Lambda$ is "ever-present". It predicts further that $p_\Lambda \neq -p_\Lambda$ ($w \neq -1$) and that $\Lambda$ has probably changed its sign many times in the past. The model contains a single free parameter of order unity that must be neither too big nor too small. In principle the value of this parameter is calculable, but for now it can only be set by hand. In this connection, it's intriguing that there exists an analog condensed matter system the "fluid membrane", whose analogous parameter is not only calculable in principle from known physics, but might also be measurable in the laboratory! That our model so far presupposes spatial homogeneity and isotropy is no doubt its weakest feature. Indeed, the ansatz on which it is based strongly suggests a generalization such that $\Lambda$-fluctuations in "causally disconnected" regions would be independent of each other; and in such a generalization, spatial inhomogeneities would inevitably arise.

Another problem is what Freeman Dyson characterizes as the struggle between order and entropy in the big crunch. As the universe approaches infinity and the average density approaches zero, temperature does not approach zero, and thus the nature of the struggle between order and entropy may actually be characterized by very different time evolutions that those with which we are familiar. In addition, there is the "Maxwell's Demon" family of arguments. This is a systems dynamic which is particularly relevant to complexity science. The problem here is that there may be emergent phenomena at the end of the life of the universe which causes the system's time evolution to then behave in unexpected ways. In some sense this is logical trap lurking behind statistical reasoning. Under normal conditions, the descriptive and inferential statistical conjecture that the near future will look like the recent past (or more boldly that fundamental physical constants are perfect invariants) is entirely reasonable. However, in the face of emergent phenomena, this assumption may not hold. Indeed, this problem is at the center of much of the debate over "relic" radiation and arguments over the age of the universe.

Yet another problem, which may also encompass emergent behavior, has to do with symmetry breaking. The universe has undergone several phase transitions by symmetry breaking. As a result, additional forces have emerged at each of these transitions. First gravity separated out of other forces, and it is for that reason that we can expect gravity wave detectors to probe more deeply into the early history of the universe than any other technology. To return to the emergent properties argument, we cannot definitively rule out (by means of present theory and observations) the possibility that at some future time (presumably near the end of the system's time evolution) that some fifth force will separate out from the known four forces.[6] At the classical level, time reversibility and a thermodynamic arrow of time is no longer problematical, but at the quantum level, and at the cosmological level, the concept remains murky at best.

---

[5] A substantial amount of work on non-extensive statistical mechanics has been done by Tsallis, et al. most recently (2007) "Nonergodicity and Central Limit Behavior for Long-range Hamiltonians" http://arxiv.org/PS_cache/arxiv/pdf/0706/0706.4021v3.pdf

## 5   Lynds' Conjecture

Peter Lynds has developed an alternative cosmology, or an alternative foundation for cosmology which flows in part from his treatment of time. He introduces his approach by stating:[7]

> Based on the conjecture that rather than the second law of thermodynamics inevitably be breached as matter approaches a big crunch or a black hole singularity, the order of events should reverse, a model of the universe that resolves a number of longstanding problems and paradoxes in cosmology is presented. A universe that has no beginning (and no need for one), no ending, but yet is finite, is without singularities, precludes time travel, in which events are neither determined by initial or final conditions, and problems such as why the universe has a low entropy past, or conditions at the big bang appear to be so special, require no causal explanation, is the result. This model also has some profound philosophical implications.

The model arises in part as a consequence of Lynds' unique treatment of time, and his ability to present a scientific framework which dispenses with the conventional notion of "instants" and a concomitant "flow" of these instants of time. He develops his cosmology based on the conjecture that in a "big crunch", at precisely the moment where the second law of thermodynamics would necessarily be breached in order to preserve symmetrical event structure, and just before the universe reaches a singularity, instead of breaching the second law of thermodynamics, the order of events would be reversed and universal expansion would begin without a singularity actually having been reached.  In Lynds words:[8]

> The natural question then became, what would happen if the second law of thermodynamics were breached? People such as Hawking (1996, 1999) and Gold had assumed that all physical processes would go into reverse. In other words, they had assumed that events would take place in the direction in which entropy was decreasing, rather than increasing as we observe today. Furthermore, they had assumed that entropy would decrease in the direction in which the universe contracted towards a big crunch (in their case, towards what we call the big bang). But if the second law correctly holds, on a large scale, entropy should still always increase. Indeed, what marks it out so much from the other laws of physics in the first place, is that it is asymmetric – it is not reversible. If all of the laws of physics, with the exception of the second law of thermodynamics, are time symmetric and can equally be reversed, it became apparent that if faced with a situation where entropy might be forced to decrease rather than increase, rather than actually doing so, the order of events should simply reverse, so that the order in which they took place would still be in the direction in which entropy was increasing. The second law would continue to hold, events would remain continuous, and no other law of physics would be contravened.

---

[6] Admittedly, this is a significant part of the epistemological argument put forth by Carroll and Carroll and Chen in the "natural" universe conjecture.  Implicit in their theory is the idea that if cosmos formation is a "natural" phenomena, rather than the "unnatural" situation suggested by the differences in scale values for fundamental constants, then an additional, emergent force would be "unnatural".

[7] Lynds, P. (2006) "On a finite universe with no beginning or end", http://arxiv.org/ftp/physics/papers/0612/0612053.pdf

[8] Ibid.

Hence, in Lynds' model, any events which would have taken place in a situation where entropy was decreasing would experience a reversal of the time ordering of events, and in the subsequent expansion of the universe, "events would immediately take up at where the big crunch singularity would have been had events not reversed, and in this direction, no singularity would be encountered. The universe would then expand from where the big crunch singularity would have been had events not reversed (i.e. the big crunch reversed), and with events going in this direction, entropy would still be increasing, no singularity would be encountered, and no laws of physics would be contravened. They would all still hold." (p. 6)

Lynds' argument necessarily bounds this reversal in the ordering of events to a very small region, and quite shortly thereafter, normal processes of inflation, including increasing entropy resume. Both the physical and the philosophical implications of this position are profound. On the philosophical side, Lynds has introduced a new concept, not only of the ultimate origin of the universe, but also a complex redefinition of "past" and "future":

> At this point, it becomes apparent that this would not only lead things back to the big bang, but it would actually cause it. The universe would then expand, cool, and eventually our solar system would take shape. It would also mean that this would be the exact repeat of the universe we live in now. Something further becomes evident, however, and it is perhaps the most important (and will probably be the most misunderstood and puzzled over) feature of this model. If one asks the question, what caused the big bang? The answer here is the big crunch. This is strange enough. But is the big crunch in the past or the future of the big bang? It could equally be said to be either. Likewise, is the big bang in the past or future of the big crunch? Again, it could equally be said to be either. The differentiation between past and future becomes completely meaningless. Moreover, one is now faced with a universe that has neither a beginning nor end

> in time, but yet is also finite and needs no beginning. The finite vs. infinite paradox of Kant completely disappears.
>
> Although if viewed from our normal conception of past and future (where we make a differentiation), the universe would repeat over and over an infinite number of times, and could also be said to have done so in the past. Crucially, however, if one thinks about what is actually happening in respect to time, no universe is in the future or past of another one. It is exactly the same version, once, and it is non-cyclic. If so desired, one might also picture the situation as an infinite number of the same universe repeating at exactly the same time. But again, if properly taking into account what is happening in respect to time, in actuality, there is no infinite number of universes. It is one and the same.

As previously indicated, this conjecture represents another radical and novel interpretation of time. However, one of the most interesting features of Lynds' conjecture is that it actually meets the two primary criteria of Hawking's M-Theory, (a) weak singularity (in Lynds' case the singularity is there, but it is, in some sense, outside the light cone and outside the observable event horizon) and (b) no boundary condition (albeit, not in the precise fashion that Hawking interprets the no boundary condition restriction).[9] Admittedly, the model is in some ways profoundly counter-intuitive, but that is largely because in a very curious way, even when treating subjects in both relativity and quantum mechanics, we have a tendency to either overtly or covertly introduce Newtonian notions of time. Some of this is addressed in Smolin's critiques of general relativity and quantum mechanics.[10] Lynds himself addresses a potential source of difficulty in the section of his article entitled "Potential Criticisms": (p.9)

---

[9] It is important to note that when the clock restarts at the big bang, the universe is not in the future or past of another one. In a sense, it is time itself that restarts (although, again, nothing in fact actually "restarts"), so there is no past or future universe. Because of this, no conservation laws are violated. It is also important to note that it is simply just the order of events that reverse - something that would be immediate. Time does not begin "flowing" backwards to the big bang, nor does anything travel into the future or past of anything, including time and some imagined "present moment". Indeed, this model contains another interesting consequence. As there is no differentiation between past and future in it, and, strictly speaking, no event could ever be said to be in the future or past of another one, it would appear to provide a clear reason as to why time travel is not possible. In relation to future and past, there is clearly nothing there to travel into. Physically speaking, the same can be said for travel through an interval of time, a flow of it, as well as space-time. (p. 8)

[10] In "Three Roads to Quantum Gravity", Smolin argues that the fundamental flaw in relativity is that it fails to incorporate the effect of the observer on observed phenomena and that quantum mechanics, while achieving the former has a tendency to treat quantum-mechanical events as occurring in traditional, Newtonian spacetime. He then argues that the unification of the partial completeness of these two new physical paradigms will be required to develop an adequate theory of quantum gravity. A quantum cosmology is likewise implicit in such a unification. Lynds provides some interesting clues to this unification insofar as puts time on all scales on a firm Einsteinian footing. In fact, one might answer Smolin's provocative essay title "Where are the Einsteinians?" with the retort "In New Zealand".

An obvious criticism for the model seems to raise itself. It implies that the universe can somehow anticipate future or past events in exact detail, and then play them over at will. At first glance, this just seems too far-fetched. How could it possibly *know*? With a little more thought, however, one recognizes that such a contention would assume that there actually was a differentiation between past and future events in the universe. With this model, it is clear there would not be. Events could neither be said to be in the future or past of one another; they would just be. Moreover, as there is nothing to make one time (as indicated by a clock) any more special than another, there is no present moment gradually unfolding; all different events and times share equal reality (in respect to time, none except for the interval used as the reference). Although physical continuity (i.e. the capability for events to be continuous), and as such, the capability for motion and of clocks and rulers to represent intervals, would stop them from all happening at once (and to happen at all), all events and times in the universe would already be mapped out. As such, as long as it still obeyed all of its own physical laws, the universe would be free and able to play any order of events it wished. Please note that this timeless picture of reality is actually the same as that provided by relativity and the "block" universe model, the formalized view of space-time resulting from the lack of a "preferred" present moment in Einstein's relativity theories, in which all times and events in the universe – past, present and future – are all mapped out together, fixed, and share equal status.

Lynds model contains a number of additional features, including some novel treatments of Kaon decay, black holes and "white holes", all of which are successfully incorporated in his model. While it is beyond the scope of the present paper to discuss these details, they deserve mention as indicators of the level of sophistication in what some might initially imagine to be a naïve interpretation of quantum cosmology.

## 6 Conclusion

In the foregoing paper we have examined a number of cosmological dynamics, particularly in light of the recent theories of time and cosmology put forward by Peter Lynds. Further, we have noted how the connection between the necessary conditions of extendibility and incompleteness with respect to Einstein's field equations leads under a variety of conditions to "unobservable singularity". The novelty of Lynds' solution is that it suggests that while the primordial singularity, including the problematic initial conditions of "specialness" explained by Carroll and Chen, exists within the light cone, it is nonetheless an inaccessible geodesic. Further, Lynds argument offers the novelty of a closed causal loop between the Big Bang and Big Crunch, no longer requiring an explanation for the special or natural entropic and scale conditions of the observable universe.